%% file: main.tex
\pdfoutput=1
\documentclass[runningheads]{llncs}

\usepackage{graphicx}
\usepackage{booktabs}
\usepackage{amsmath}
\usepackage{multirow}
\usepackage{tikz}
\usetikzlibrary{arrows.meta}
\usepackage{hyperref}
\usepackage{etoolbox}
\usepackage{eso-pic}
\AtBeginEnvironment{thebibliography}{\footnotesize}

\newcommand\copyrighttext{%
\footnotesize
This manuscript has been accepted for presentation at the
52nd Euromicro Conference on Software Engineering and Advanced Applications (SEAA) 2026,
held in Krakow, Poland, 02 -- 04 September 2026, and for publication
in Springer Lecture Notes in Computer Science (LNCS) proceedings.
This is the author's accepted manuscript version.
The final authenticated publication will be available via Springer.
}
\newcommand{\copyrightnotice}{%
\AddToShipoutPictureFG*{%
\AtPageLowerLeft{%
\raisebox{1cm}{%
\makebox[\paperwidth][c]{%
\fbox{\parbox[b]{0.9\textwidth}{\copyrighttext}}%
}}}}}

\graphicspath{{figures/}}

\begin{document}

\copyrightnotice

\title{Mining Architectural Quality Under Agentic AI Adoption: A Causal Study of Java Repositories}
\titlerunning{Agentic AI and Architectural Quality}
\authorrunning{O. A. Larsen and M. T. Moghaddam}

\author{Oliver Aleksander Larsen\inst{1}\thanks{Corresponding author.} \and Mahyar T. Moghaddam\inst{1}}
\institute{SDU Software Engineering, University of Southern Denmark,\\
Odense, Denmark\\
\email{\{olar,mtmo\}@mmmi.sdu.dk}}

\maketitle

\begin{abstract}
AI coding tools are now used by a majority of developers, and agentic
use of these tools has popularized the practice colloquially called
``vibe coding''. Yet causal evidence on their effect on software
architecture is scarce. Prior
causal work has measured code-level outcomes (complexity, static
analysis warnings); whether such degradation propagates to
architecture-level outcomes remains unknown. We mine 151 open-source
Java repositories, 74 with detectable agentic AI adoption (identified
via configuration files and \texttt{Co-Authored-By} commit trailers)
and 77 propensity-matched controls, across a 13-month per-repository
window yielding 1,811 monthly Arcan snapshots. We estimate the causal
effect of adoption on architectural smell density (ASD) with a
staggered difference-in-differences design and the Borusyak imputation
estimator, applying a causal design recently used for code-level
metrics to the architecture level. Total smell counts are essentially
unchanged ($+1.1\%$, $p = 0.82$) while lines of code grow $+12.8\%$
($p = 0.003$); the resulting 6.7\% ASD decline ($p = 0.004$) is
therefore a denominator effect rather than an architectural
improvement. Per-type estimates and robustness checks (wild cluster
bootstrap, Lee bounds, stale-observation sensitivity) corroborate the
pattern; pre-trends are flat (Wald $p = 0.90$), consistent with
parallel trends. Density-normalized outcomes can mislead
when treatment affects system size: raw counts and explicit
decomposition are required for causal mining studies of AI tool
adoption. The complete replication package, including the curated
151-repository monthly panel, is publicly available.
\end{abstract}

\keywords{architectural smells \and architectural erosion \and AI-assisted development \and vibe coding \and mining software repositories \and difference-in-differences}

\input{sections/01_introduction}
\input{sections/02_background_and_related_work}
\input{sections/03_methodology}

\input{sections/04_results}

\input{sections/05_discussion}
\input{sections/06_threats}
\input{sections/07_conclusion}

\section*{Data Availability}
\begin{sloppypar}
The replication package (pipeline scripts, processed monthly panel,
configurations, fixed seeds, Docker setup) is archived on Zenodo
(\url{https://doi.org/10.5281/zenodo.20510047}) and mirrored at
\url{https://github.com/Oliver1703dk/seaa2026-replication-package}; all
tables and figures reproduce via \texttt{make analyze}.
\end{sloppypar}

\bibliographystyle{splncs04}
\bibliography{references}

\end{document}

%% file: sections/01_introduction.tex
\section{Introduction}\label{sec:introduction}

AI coding tools have moved from inline autocomplete to fully agentic
operation: given a natural-language intent, tools such as Cursor,
GitHub Copilot, Claude Code, and Aider generate, edit, and refactor
code across multiple files~\cite{github2025copilot,cursor2024,anthropic2025claudecode}.
The resulting paradigm, popularized as ``vibe coding'' by
Karpathy~\cite{karpathy2025vibe}, leaves observable artifacts in
repositories: tool-specific configuration files, which encode
project-specific directives consistent with agentic
engagement~\cite{jiang2026beyond}, and \texttt{Co-Authored-By}
commit trailers; neither signal arises from inline autocomplete. These artifacts make agentic
adoption visible to repository mining at scale, even as practitioner
surveys report concerns about long-term
maintainability~\cite{waseem2025vibe,fawzy2026vibe}.

These concerns have empirical support at the code level. He et
al.~\cite{he2026speed} studied 806 Cursor-adopting repositories using
a staggered difference-in-differences (DiD) design and found that AI
adoption increases static analysis warnings by 30\% and code
complexity by 41\%. Yet code-level complexity and architectural
quality are distinct: a method with high cyclomatic complexity does
not necessarily create a cyclic dependency; a duplicated code block
does not produce a hub-like component. Esposito et
al.~\cite{esposito2025correlation} showed that 33.8\% of static
analysis warnings never co-occur with any architectural smell,
confirming that code-level and architectural degradation are only
partially overlapping concerns. Whether code-level degradation
propagates to architecture-level degradation remains unanswered.

The gap matters. Architectural smells are recurring structural
anti-patterns in a system's dependency
graph~\cite{garcia2009toward} and are indicators of architectural
erosion~\cite{vangurp2002design,hochstein2005combating} that affects
maintainability and
evolvability~\cite{jolak2025maintainability,fontana2022evolution}.
Qualitative work suggests the risk is
real~\cite{amasanti2025impact}, yet no study has measured
architectural impact with a causal mining design.

The code-level evidence above predicts that agentic AI adoption
should increase architectural smell density (ASD), particularly
structural anti-patterns plausibly arising from file-scoped
generation. RQ1--RQ3 test that prediction at the architectural
level; RQ4 decomposes any observed ASD effect into numerator and
denominator channels.\smallskip

\noindent\textbf{RQ1:} Does the adoption of agentic AI coding tools affect architectural
smell density (ASD) in open-source Java repositories?\\[2pt]
\textbf{RQ2:} Which categories of architectural smells show the largest treatment
effect?\\[2pt]
\textbf{RQ3:} How does the effect on ASD evolve over time after agentic AI tool adoption?\\[2pt]
\textbf{RQ4:} Is the observed ASD effect driven by changes in architectural
smell counts (numerator), code volume (denominator), or both?\smallskip

We mine 151 open-source Java repositories, 74 with observable
agentic AI adoption (detected via configuration files and
\texttt{Co-Authored-By} trailers) and 77 propensity-matched
controls~\cite{rubin1980bias,stuart2010matching}, yielding 1{,}811 monthly Arcan
snapshots~\cite{fontana2017arcan,sas2023atd} over 13 months per
repository. The causal effect on ASD is estimated with a staggered
difference-in-differences design and the Borusyak imputation
estimator~\cite{borusyak2024revisiting}, applied here at the
architectural level (one abstraction higher than He et
al.~\cite{he2026speed}).

ASD declines by 6.7\% ($p = 0.004$), the opposite-direction finding
to that prediction. Decomposition reveals the mechanism: total smell
counts are essentially unchanged ($+1.1\%$, $p = 0.82$) while code
volume grows substantially ($+12.8\%$, $p = 0.003$); the decline is
therefore a composition effect driven by the denominator. Pre-trends
are flat (Wald $p = 0.90$), consistent with parallel trends.

This paper contributes: (1)~a causal mining pipeline (151
repositories, 1{,}811 monthly snapshots, Borusyak DiD on
architectural metrics) reusable for other AI-tool quality outcomes;
(2)~a methodological warning that density-normalized outcomes
mislead when treatment affects system size, requiring explicit
numerator/denominator decomposition; (3)~the first causal evidence
on architecture-level effects of agentic AI adoption, consistent
with divergence between code- and architecture-level
effects~\cite{he2026speed,esposito2025correlation}; and (4)~a
complete open-source replication package with the curated 151-repo
monthly panel.

%% file: sections/02_background_and_related_work.tex
\section{Background and Related Work}\label{sec:background}

\subsection{AI Coding Tools and the Detection Problem in Repository Mining}\label{sec:bg_vibe}

AI tools are now used by 63.2\% of professional
developers~\cite{stackoverflow2024survey}. The major
tools (GitHub Copilot, Cursor, Claude Code) have each evolved
beyond inline autocomplete to offer agentic capabilities: given a
natural-language intent, the tool autonomously generates, edits, and
refactors code across multiple files~\cite{github2025copilot,cursor2024,anthropic2025claudecode}.
Practitioner accounts describe developers accepting AI-generated
code with minimal structural
review~\cite{karpathy2025vibe,fawzy2026vibe,waseem2025vibe}.

The key distinction for repository mining is between \emph{usage
modes}, not between tools. In autocomplete mode, the developer
accepts inline suggestions; no project-level configuration is needed
and no attribution metadata is produced. In agentic mode, developers
create tool-specific configuration files (which encode
project-specific directives~\cite{jiang2026beyond}) and produce
\texttt{Co-Authored-By} commit trailers; neither artifact arises
from autocomplete-only usage. Any
detection strategy based on repository artifacts (including ours and
He et al.'s~\cite{he2026speed}) therefore observes only agentic
usage (Sec.~\ref{sec:meth_sample}). This is also the mode posing the
greatest architectural risk: agentic tools operate at file and
function scope, while architecture is an emergent property of
inter-module relationships~\cite{bass2021software}.

\subsection{Architectural Erosion and Architectural Smells}\label{sec:bg_erosion}

Perry and Wolf~\cite{perry1992foundations} formalized software
architecture as the set of design decisions governing a system's
structural organization. \emph{Architectural erosion} is the gradual
divergence between intended and implemented
architecture~\cite{vangurp2002design,hochstein2005combating}.
\emph{Architectural smells} operationalize erosion as detectable
structural anti-patterns in dependency
graphs~\cite{garcia2009toward}. We measure four types, each
violating a known design
principle~\cite{martin2003agile,lippert2006refactoring}:

\begin{itemize}\setlength{\itemsep}{1pt}
\item \textbf{Cyclic Dependency (CD):} circular dependency chains among
  packages, violating the Acyclic Dependencies Principle.
\item \textbf{Unstable Dependency (UD):} a package depending on a
  less-stable package, violating the Stable Dependencies Principle.
\item \textbf{Hub-Like Dependency (HL):} a component with both high fan-in
  and high fan-out, undermining modularity.
\item \textbf{God Component (GC):} an oversized package concentrating too
  many responsibilities, violating modularity.
\end{itemize}

\noindent We detect these smells with
Arcan~\cite{fontana2017arcan}, with reported precision in the
$70$--$100\%$ range across Java projects~\cite{fontana2017arcan,sas2023atd}.
Architectural smell density
(ASD\,=\,total smells\,/\,KLOC) is used as a proxy for the rate of
erosion; erosion proper requires comparison against an intended
architecture~\cite{vangurp2002design}, which is unavailable for
open-source projects at scale. ASD requires decomposition when
treatment affects code volume (Sec.~\ref{sec:res_decomposition}).

\subsection{Empirical Studies of AI-Assisted Development}\label{sec:rw_ai}

He et al.~\cite{he2026speed} conducted the closest predecessor study:
a staggered DiD analysis of 806 Cursor-adopting repositories,
reporting $+30\%$ static-analysis warnings and $+41\%$ complexity
alongside higher development velocity. We follow their mining design
one abstraction level higher: from SonarQube's intra-method metrics
to Arcan's inter-package metrics. Esposito et
al.~\cite{esposito2025correlation} confirmed the gap between these
two levels: 33.8\% of static analysis warnings are architecturally
benign, unassociated with any smell, indicating that code-level
findings cannot be assumed to translate to architecture-level effects.

Beyond causal studies, Cotroneo et al.~\cite{cotroneo2025human}
found AI-generated code more prone to vulnerabilities, while
Yeti\c{s}tiren et al.~\cite{yetistiren2023evaluating} report mixed
quality results across major AI tools; both are cross-sectional and
measure code-level properties exclusively. Amasanti and Jahi\'c~\cite{amasanti2025impact} survey practitioners
about architectural-quality concerns under AI assistance, finding
mixed signals; Waseem et al.~\cite{waseem2025vibe} document related
qualitative concerns.
Jiang and Nam~\cite{jiang2026beyond} characterized Cursor
configuration files across 401 repositories, finding
project-specific directives consistent with agentic engagement. A
companion study by Agarwal et al.~\cite{agarwal2026agents} extends
He et al.'s analysis by distinguishing IDE-based AI assistants from
autonomous coding agents, finding larger quality effects for
agentic usage. These studies establish a code-level signal but do
not address the architectural channel; the present work fills that
gap.

\subsection{Architectural Smell Evolution}\label{sec:rw_erosion}

Longitudinal studies of smell evolution provide the baseline against
which AI-induced changes must be assessed. Sas et
al.~\cite{fontana2022evolution} found that smells persist, compound,
and resist removal: erosion is the natural trajectory of organic
development. Gnoyke et al.~\cite{gnoyke2024evolution} confirmed
these patterns across 485 releases of 14 open-source systems. Our
DiD design controls for this natural trajectory by measuring the
\emph{differential} change in treated repositories relative to
controls. Related work links smells to reduced
maintainability~\cite{jolak2025maintainability}, validates Arcan
against LLM-based detection~\cite{tessa2025smells}, and identifies
conformance checking as
underexplored~\cite{schmid2025slr}.

\textbf{Summary.} No prior work applies a causal mining design to
measure the architectural impact of AI-assisted development. The
closest causal precedent (He et al.) operates on code-level metrics;
the closest architectural-level work (Amasanti and Jahi\'c) is
qualitative.

%% file: sections/03_methodology.tex
\section{Mining Design and Statistical Model}\label{sec:methodology}

\input{figures/fig_pipeline}

\subsection{Study Design}\label{sec:meth_design}

We employ a staggered difference-in-differences (DiD) design to
estimate the causal effect of observable agentic AI adoption on
architectural quality. Figure~\ref{fig:pipeline} summarizes the four phases
(mining, matching, extraction, analysis) whose sample sizes we trace below. Treatment is the first detectable adoption of
an agentic AI coding tool (Sec.~\ref{sec:meth_sample}); controls are
matched non-adopters observed over the same calendar periods. Each
repository contributes a 13-month window (six pre-adoption, adoption
month, six post-adoption). The unit of analysis is the
repository-month; the primary outcome is
$\log(\text{ASD} + 1)$ with
$\text{ASD} = \text{total\_smells}/\text{KLOC}$ (thousand lines of
code); effects are reported as $100 \times (\exp(\hat{\beta}) - 1)\%$
(approximate under $\log(x+1)$; see Sec.~\ref{sec:res_decomposition}
for the strict-log specification on positive observations).
Identification relies on
repository and time fixed effects under parallel trends. Repository
sampling follows the methodological caveats of Kalliamvakou et
al.~\cite{kalliamvakou2016promises} (forks and inactive
repositories filtered).

\subsection{Treatment Detection and Sample Construction}\label{sec:meth_sample}

\begin{sloppypar}
\noindent\textbf{Treatment identification.}
AI tool adoption is identified through two categories of repository
artifacts.
First, tool-specific configuration files per tool:
Cursor (\texttt{.cursorrules}, \texttt{.cursor/rules/}),
Copilot (\texttt{.github/copilot-instructions.md}),
Claude Code (\texttt{CLAUDE.md} or \texttt{claude.md}),
Aider (\texttt{.aider.conf.yml}, \texttt{.aider.model.settings.yml}),
and Codex (\texttt{AGENTS.md}).
Second, \texttt{Co-Authored-By} commit trailers matching
known AI tool patterns.
\end{sloppypar}

\paragraph{Detection patterns.}
Each configuration marker is matched as a literal pathspec via
\texttt{git log -{}-all -{}-diff-filter=A}; all markers are file pathspecs
except \texttt{.cursor/rules/}, which is a directory pathspec
(any addition under it counts). Commit-message signals are matched via
\texttt{git log -{}-all -E -{}-grep} against four POSIX-ERE patterns: three
case-insensitive \texttt{Co-Authored-By:} trailers naming
\texttt{Claude}, \texttt{aider}, or \texttt{Copilot}, plus the
literal subject-line tag \texttt{(aider)} that the Aider CLI appends
to commits it authors. The adoption date is the earliest event
across both signal types. Ambiguous markers (\texttt{CLAUDE.md}/\texttt{claude.md},
\texttt{AGENTS.md}) are content-validated, while high-reliability markers
(\texttt{.cursorrules}, \texttt{copilot-instructions.md},
\texttt{.aider.*}) are accepted directly. Full patterns and code are
in the replication package.

Both categories indicate configured, agentic usage rather than
inline autocomplete, which produces no detectable artifacts
(Sec.~\ref{sec:bg_vibe}). The construct measured is observable
agentic AI adoption, distinct from the broader behavioral pattern of
accepting AI output without review. Undetected autocomplete in
controls induces attenuation bias, making estimates conservative.

\paragraph{Sensitivity to autocomplete contamination.}
If a fraction~$c$ of controls use undetected AI tools, attenuation
bias yields $\hat{\beta}_{\text{true}} \approx
\hat{\beta}_{\text{obs}} / (1 - c)$~\cite{aigner1973regression};
the corrected effect at the empirical $c = 8.9\%$ is $-7.4\%$,
amplifying rather than altering the neutrality conclusion.

Tool composition (matched / post-Arcan retained): Claude Code 52/44
(one of which also uses Aider), Copilot 21/14, Cursor 16/15, Codex
1/1. Single-event sensitivity is in Sec.~\ref{sec:res_robustness};
tool-stratified analysis in Sec.~\ref{sec:res_tool_heterogeneity}. We validate the proxy
empirically via temporal alignment: LOC growth emerges at adoption
dates (Sec.~\ref{sec:res_decomposition}).

\noindent\textbf{Inclusion criteria.}
Treatment candidates satisfy:
(1) Java as primary language;
(2) GitHub repository size $\geq$2{,}000 KB at mining time as a
code-volume proxy (LOC and package counts are post-hoc Arcan outputs);
(3) $\geq$2 commits per month on average;
(4) $\geq$6 months pre-adoption history; and
(5) $\geq$3 months post-adoption history. These criteria yield 90
treatment repositories from 204 candidates
(Fig.~\ref{fig:pipeline}).

\noindent\textbf{Control matching.}
The control pool comprises 580 Java repositories without detectable
AI markers. We
perform 1:1 nearest-neighbor matching~\cite{rubin1980bias,stuart2010matching}
on Euclidean distance over six standardized covariates (four
log-transformed to address skew: stars, forks, commits, size), with
a propensity-score caliper of
$0.2\times\text{SD}(\text{logit}(\hat{p}))$~\cite{rosenbaum1985constructing,austin2011caliper}.
Matching yields 90 pairs (the 90+90 in Fig.~\ref{fig:pipeline}). Post-matching,
four of six covariates satisfy $|\text{SMD}| < 0.1$ (standardized mean
difference; below 0.1 is well balanced); repository size
($\text{SMD} = -0.18$) and commit count ($\text{SMD} = -0.12$) exceed it,
addressed via covariate adjustment in Sec.~\ref{sec:res_robustness}.

\subsection{Architectural Smell Extraction}\label{sec:meth_extraction}

\noindent\textbf{Monthly snapshots and Arcan analysis.}
For each repository-month we extract the last commit before the
first of the month (\texttt{git rev-list --first-parent --before})
and materialize it as a temporary working copy. Of the 180 matched
repositories (90 treatment, 90 control), 5 controls are dropped before
analysis for having no resolvable commit in their study window
(insufficient git history), leaving 175 repositories (90 treatment, 85
control) that yield 2,241 monthly snapshots. Each snapshot is analyzed with
Arcan 2 CLI~\cite{fontana2017arcan} (trial edition, pinned Docker image)
detecting the four smell types from Sec.~\ref{sec:bg_erosion};
analysis succeeded for 1,811 of 2,241 snapshots (80.8\%, failures
from JVM memory limits or incompatible structures). A further 24
repositories with zero successful snapshots (16 treatment, 8
control) are excluded, yielding the final panel of 1,811 observations
across 151 repositories (74 treatment, 77 control). Total attrition is thus
29 of 180 matched repositories (16.1\%; Sec.~\ref{sec:threats}).

\noindent\textbf{Metrics.}
Primary outcome: $\text{ASD} = \text{total\_smells}/\text{KLOC}$,
aggregated across class and package granularity (a container-only
sensitivity is in Sec.~\ref{sec:res_robustness}). Cyclic
dependencies dominate (mean CD\,=\,333 vs.\ $<$10 for other
types), so we complement aggregate ASD with per-type analysis (RQ2),
balanced-ASD, and total-count decomposition. Secondary metrics:
efferent coupling ($Ce$), instability ($I = Ce/(Ca + Ce)$), and
distance from the main sequence. All outcomes log-transformed as
$\log(x + 1)$. The primary specification uses fixed effects only;
the secondary specification, two-way fixed effects with the Sun and
Abraham~\cite{sun2021estimating} interaction-weighted correction
(TWFE-SA), additionally adjusts for monthly contributors.

\subsection{Statistical Model}\label{sec:meth_model}

Because repositories adopt at different calendar months (staggered
adoption), standard two-way fixed effects estimation can produce
biased estimates under heterogeneous treatment
effects~\cite{goodman2021difference,dechaisemartin2020two}. We
therefore employ estimators designed for staggered settings, and adopt
Borusyak imputation as primary because, unlike standard TWFE, it never uses
already-treated repositories as controls and so stays unbiased under
heterogeneous treatment effects.

\noindent\textbf{Primary estimator: Borusyak imputation.}
Following Borusyak et al.~\cite{borusyak2024revisiting}:
\begin{equation}\label{eq:did}
\log(\text{ASD}_{it} + 1) = \alpha_i + \gamma_t + \beta \cdot D_{it}
  + \varepsilon_{it}
\end{equation}
where $i$ indexes repositories, $t$ months, $\alpha_i$ and
$\gamma_t$ are repository and time fixed effects, and $D_{it} = 1$
once repository $i$ has adopted. The imputation procedure fits on
untreated observations (controls plus pre-adoption periods of
treated), predicts counterfactual $\hat{Y}_{it}(0)$ for treated
post-adoption observations, and estimates the average treatment effect on
the treated (ATT) as
$\widehat{\text{ATT}} = \overline{Y_{it} - \hat{Y}_{it}(0)}$.
Standard errors are clustered at the repository level.

Architectural metrics exhibit high inertia (intraclass correlation
ICC\,=\,0.979); fixed effects absorb between-repository variance,
isolating within-repo treatment effects. GC nulls should be
interpreted as potentially underpowered (77.5\% zero-inflation).

\noindent\textbf{Secondary estimator and event study.}
The Sun and Abraham~\cite{sun2021estimating} interaction-weighted
estimator provides an independent concordance check. Event-study coefficients
are estimated for $h \in \{-6, \ldots, +6\}$ with $h = -1$ as
reference; a Wald test on $h = -6$ through $h = -2$ jointly zero is
the parallel-trends falsification check. Post-adoption coefficients
trace the trajectory (RQ3).

\noindent\textbf{Denominator decomposition (RQ4).}
Since ASD is a ratio ($\text{smells}/\text{KLOC}$), a treatment
effect can reflect changes in the numerator, the denominator, or
both. We run separate DiD models on
$\log(\text{total\_smells}+1)$ and $\log(\text{KLOC}+1)$ to
identify the mechanism.

\noindent\textbf{Robustness checks.}
Pre-registered: a placebo test at a fake adoption date and an alternative
matching specification (subsumed by Mahalanobis with regression
adjustment). Added post-results: wild cluster
bootstrap~\cite{cameron2008bootstrap,webb2023wild} (few-cluster
robustness); Lee~\cite{lee2009training} bounds (worst-case attrition);
smells per package; stale observations; Bacon decomposition (staggered-DiD
weighting); size stratification; container-only. Post-hoc: tool-type
stratification; dose-response.

%% file: figures/fig_pipeline.tex
\begin{figure}[t]
\centering
\makebox[\textwidth][c]{%
\scalebox{0.85}{%
\begin{tikzpicture}[
  box/.style={draw, rounded corners, minimum width=2.6cm, minimum height=1.1cm,
              align=center, font=\small},
  arrow/.style={-{Stealth[length=3mm]}, thick},
  label/.style={font=\scriptsize, align=center}
]
\node[box] (mine)    at (0,0)     {Mining\\[2pt]\scriptsize 204 treatment\\[-1pt]\scriptsize 580 control};
\node[box] (match)   at (4.0,0)   {Matching\\[2pt]\scriptsize 90 + 90 repos};
\node[box] (extract) at (8.0,0)   {Extraction\\[2pt]\scriptsize 175 repos\\[-1pt]\scriptsize 2,241 snapshots};
\node[box] (analyze) at (12.0,0)  {Analysis\\[2pt]\scriptsize 1,811 obs\\[-1pt]\scriptsize 151 repos};

\draw[arrow] (mine) -- (match) node[midway, above, label] {window\\[-2pt]filter};
\draw[arrow] (match) -- (extract) node[midway, above, label] {monthly\\[-2pt]snapshots};
\draw[arrow] (extract) -- (analyze) node[midway, above, label] {Arcan\\[-2pt](80.8\%)};

\node[below=0.2cm, font=\scriptsize\itshape] at (mine.south) {Phase 1};
\node[below=0.2cm, font=\scriptsize\itshape] at (match.south) {Phase 2};
\node[below=0.2cm, font=\scriptsize\itshape] at (extract.south) {Phase 3};
\node[below=0.2cm, font=\scriptsize\itshape] at (analyze.south) {Phase 4};
\end{tikzpicture}%
}%
}
\caption{Study pipeline with sample sizes at each stage.}
\label{fig:pipeline}
\end{figure}
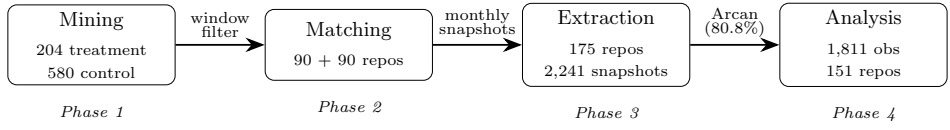

%% file: sections/04_results.tex
\section{Results}\label{sec:results}

\subsection{Sample and Descriptive Statistics}\label{sec:res_sample}

The final panel comprises 1,811 observations across 151 repositories
(74 treatment, 77 control), after excluding 29 of 180 matched repositories (16.1\%;
Sec.~\ref{sec:threats}). Groups are comparable in size (mean KLOC: 44.9
vs.\ 45.2), maturity (73.8 vs.\ 80.1 months), and stars (918 vs.\ 920).
ASD is right-skewed (median\,=\,3.0, max\,=\,58.8), motivating the log
transformation. Treatment repos exhibit higher mean ASD (6.1 vs.\ 5.1);
with ICC\,=\,0.979, repository fixed effects absorb this variation.

\subsection{RQ1: Effect of AI Adoption on Architectural Smell
  Density}\label{sec:res_rq1}

Table~\ref{tab:main_did} reports the main DiD estimates. The Borusyak
imputation estimator yields $\hat{\beta} = -0.070$
(SE\,=\,0.024, 95\% CI $[-0.117, -0.023]$, $p = 0.004$), corresponding
to 6.7\% lower ASD in repositories with observable agentic AI adoption.
Sun and Abraham confirms concordance ($-6.6\%$, $p = 0.005$).
Against the theoretical expectation of architectural degradation
(Sec.~\ref{sec:introduction}), the negative point estimate is the
opposite-direction finding; the mechanism behind the headline number
is the focus of Sec.~\ref{sec:res_decomposition}.

\input{tables/tab_main_did}

\subsection{RQ2: Disaggregated Effects by Smell Type}\label{sec:res_rq2}

\begin{figure}[t]
  \centering
  \includegraphics[width=1.00\linewidth,height=6.0cm,keepaspectratio]{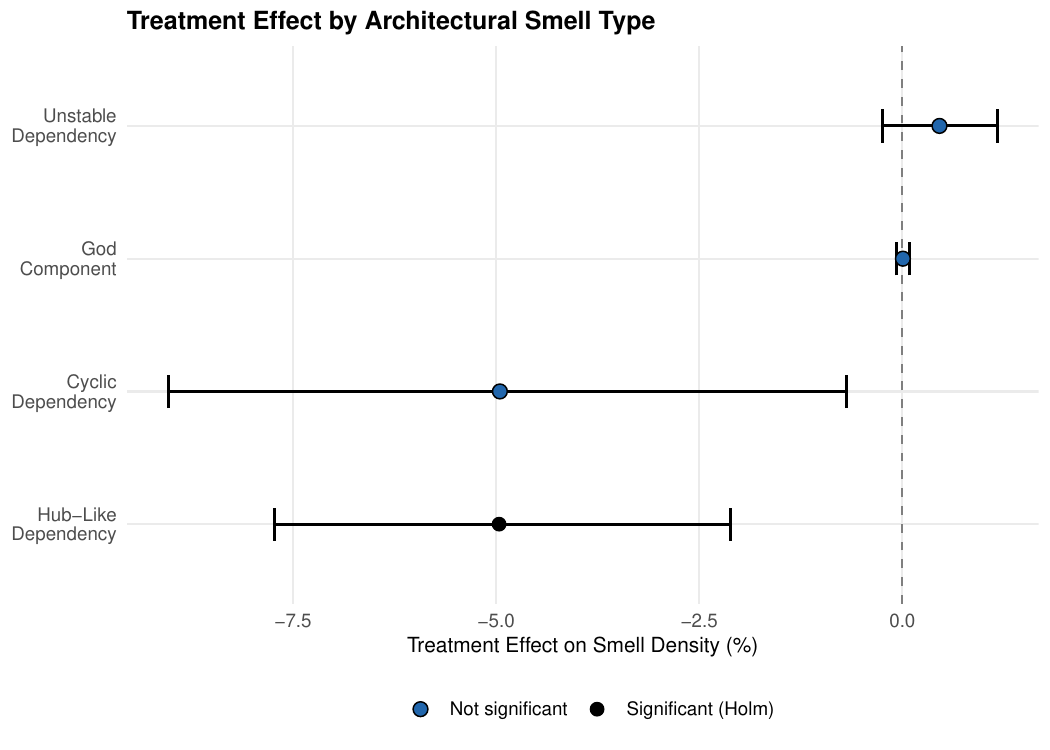}
  \caption{Per-type treatment effects (Borusyak imputation) with
  95\% CIs. Filled markers: significant after Holm correction.}
  \label{fig:rq2_forest_plot}
\end{figure}

Figure~\ref{fig:rq2_forest_plot} presents per-type treatment effects
(Borusyak imputation). Because we test four smell types, we report
Holm-adjusted $p$-values, which raise the bar so multiple tests do not
yield a false positive. Hub-like
dependency density decreases by 5.0\%
($p = 0.001$; Holm-adjusted $p = 0.003$), the only type surviving
multiplicity correction. Cyclic dependency shows a comparable estimate
($-5.0\%$, $p = 0.023$) but is non-significant after Holm adjustment
($p = 0.070$). UD and GC show no significant density change
($p = 0.205$, $p = 0.877$). The GC null likely reflects low power:
77.5\% of repo-months record zero god components.

Raw UD counts show a suggestive increase of 5.8\% (raw $p = 0.032$;
Holm-adjusted across four raw-count tests, $p = 0.128$). The density
null masks this marginal signal, discussed in
Sec.~\ref{sec:disc_neutral}. Density and count tests answer distinct
mechanistic questions, with family-wise error controlled within each
family.

\subsection{RQ3: Dynamic Treatment Effects}\label{sec:res_rq3}

\begin{figure}[t]
  \centering
  \includegraphics[width=1.00\linewidth,height=5.6cm,keepaspectratio]{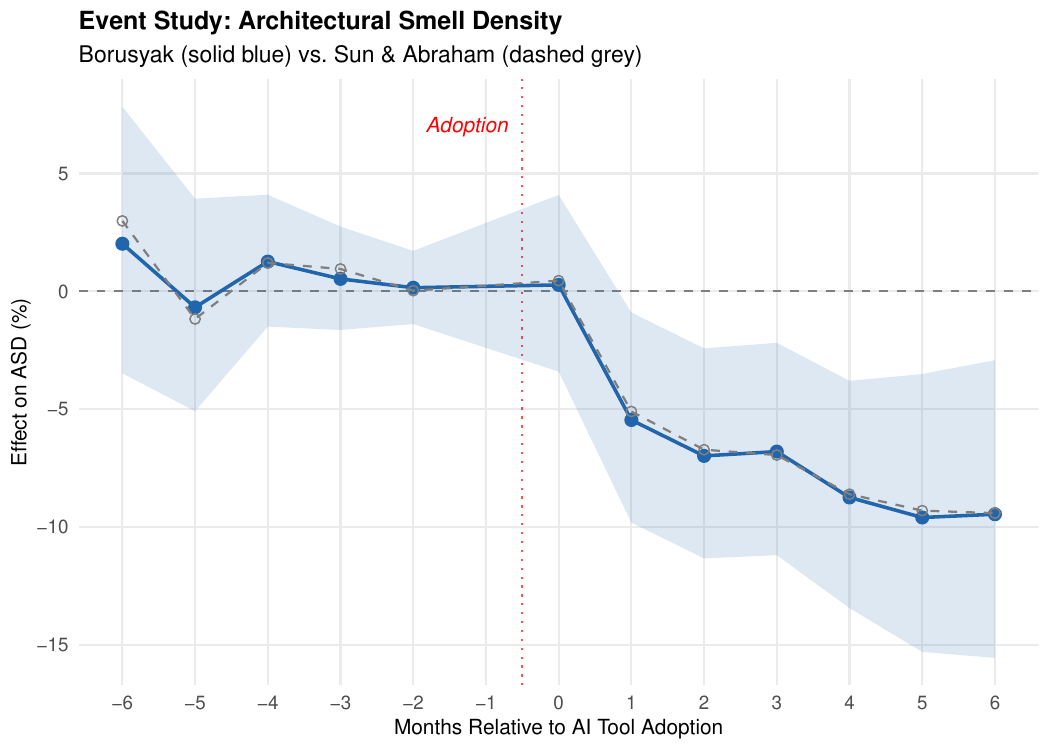}
  \caption{Event study: ASD effect by relative month. Pre-period
  coefficients near zero (Wald $p = 0.896$); post-adoption effect
  grows to ${\sim}{-9.5\%}$ at $h = 6$. Sun \& Abraham overlay
  (grey dashed) confirms concordance.}
  \label{fig:rq3_event_study}
\end{figure}

Figure~\ref{fig:rq3_event_study} presents the event study. Pre-period
coefficients ($h = -6$ to $h = -2$) cluster near zero, and the Wald
test cannot reject joint nullity ($\chi^2 = 1.64$, df\,=\,5,
$p = 0.896$), consistent with the parallel trends assumption (here a high
$p$-value is reassuring: no divergence before adoption).

Post-adoption, the effect appears immediately at $h = 1$ ($-5.5\%$)
and grows to $-9.5\%$ by $h = 6$, consistent with cumulative code
growth diluting smell density (Sec.~\ref{sec:res_decomposition}).
The effect is still expanding at $h = 6$, so longer post-adoption
windows are needed to see whether smell counts eventually grow in
step with LOC and the density gap closes.

\subsection{RQ4: Decomposing the ASD Effect}\label{sec:res_decomposition}

\begin{figure}[t]
  \centering
  \includegraphics[width=1.00\linewidth,height=7.0cm,keepaspectratio]{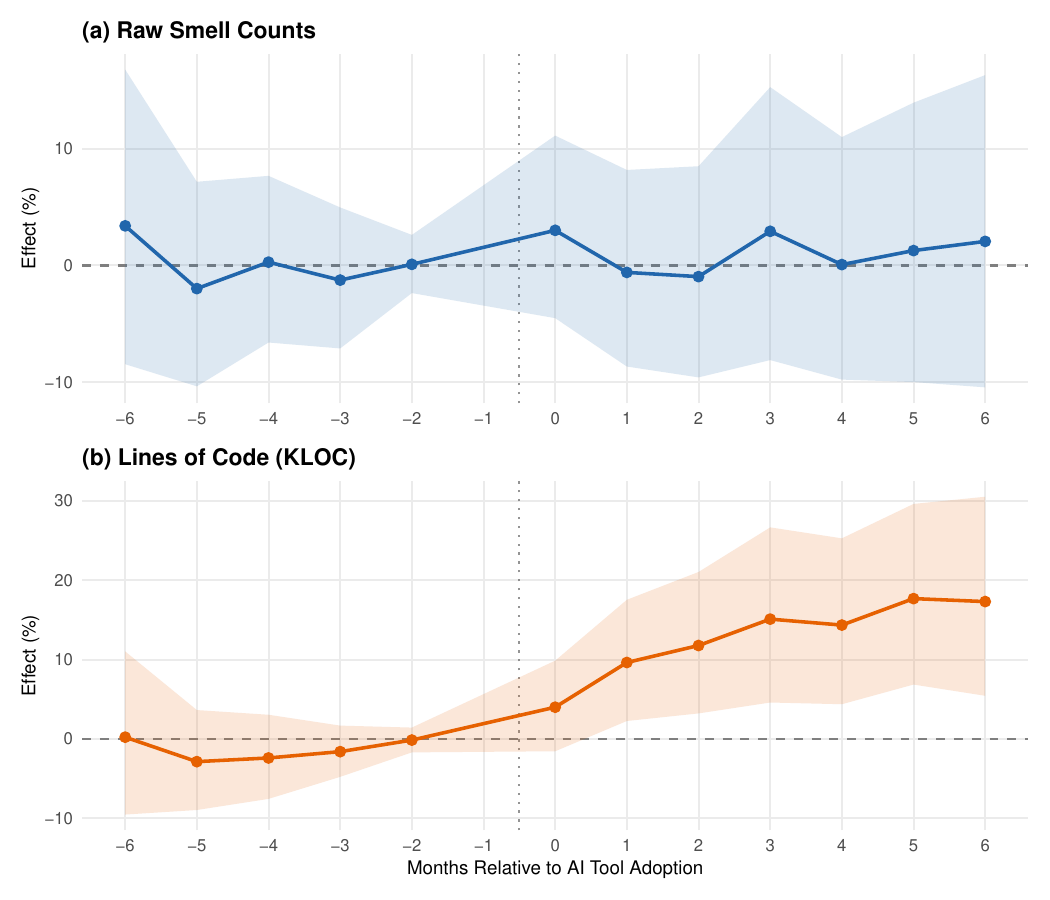}
  \caption{Decomposition: separate DiD on raw smell counts (top) and
  LOC (bottom). Smells unchanged ($+1.1\%$, $p = 0.82$); LOC grows
  significantly ($+12.8\%$, $p = 0.003$).}
  \label{fig:decomposition_panels}
\end{figure}

The 6.7\% ASD reduction could reflect fewer smells (numerator) or
faster code growth (denominator). Separate DiD models
(Fig.~\ref{fig:decomposition_panels}) reveal the mechanism. Raw smell
counts are unchanged ($+1.1\%$, $p = 0.82$) while code volume grows
significantly faster in adopters than controls ($+12.8\%$, $p = 0.003$). The ASD decrease is a
composition effect: the marginal code carries lower smell density than
the pre-existing codebase. A strict-log specification on positive
observations (which preserves the ratio identity that $\log(x+1)$
approximates) yields $\hat{\beta} = -0.139$ ($-13.0\%$, $p < 0.001$),
implying an even larger denominator share than under $\log(x+1)$.

Per-type raw counts confirm the pattern: CD, HL, and GC show no
significant absolute changes (all $p > 0.25$). UD shows a suggestive increase ($+5.8\%$, raw $p = 0.032$;
Holm-adjusted $p = 0.128$), the strongest per-type signal though
non-significant after multiplicity correction. Coupling metrics
($Ce$, $I$, $D$) are similarly stable (all $p > 0.05$).
Package count also grows ($+9.7\%$, $p = 0.032$), yet smell counts
remain unchanged despite expanding modular structure. The
methodological implication for empirical mining is discussed in
Sec.~\ref{sec:disc_decomp}.

\subsection{Exploratory: Treatment Heterogeneity by
  Tool}\label{sec:res_tool_heterogeneity}

Tool-stratified Borusyak estimates are consistently negative (Codex,
$n_T = 1$, omitted): Claude Code ($n_T = 44$, $-5.1\%$, $p = 0.039$),
Copilot ($n_T = 14$, $-4.0\%$, $p = 0.005$), and Cursor ($n_T = 15$,
$-13.5\%$, $p < 0.001$). Magnitudes differ (Cursor strongest), but with a
non-significant dose-response interaction ($p = 0.716$) and post-hoc
stratification, the larger Cursor effect is a hypothesis for replication,
not a finding.

\subsection{Robustness Checks}\label{sec:res_robustness}

Table~\ref{tab:supplementary} reports supplementary analyses.
Covariate adjustment yields $-6.6\%$ ($p = 0.004$). The Bacon
decomposition reports 89.1\% of the TWFE weight from clean
comparisons (80.2\% treated-vs-never-treated, 8.9\%
treated-vs-not-yet-treated) and 10.8\% from the potentially-biased
later-vs-already-treated comparison, which the Borusyak primary estimator
avoids by construction; a placebo at a fake adoption date ($t - 3$) is null
($p = 0.427$). The wild cluster
bootstrap~\cite{cameron2008bootstrap,webb2023wild} (small-sample-safe
significance) yields $-6.0\%$ ($p = 0.010$); Lee~\cite{lee2009training}
bounds (worst-case attrition) span $[-9.9\%, -2.3\%]$ (both negative,
though the upper-bound CI includes zero, $p = 0.239$).
Size-stratified estimates are consistent ($-6.3\%$/$-7.2\%$);
balanced ASD ($-4.6\%$, $p < 0.001$) and a multi-event restriction
excluding repositories with a single detection event ($\geq 2$ events,
$-6.8\%$, $p = 0.020$) retain significance, indicating the result is
not driven by one-off configuration commits. Container-level
(package) smells alone (an outcome restriction; $-3.0\%$, $p = 0.074$)
are directionally consistent but attenuated. Only aggressive stale
filtering ($N = 94$, $p = 0.223$) attenuates to non-significance;
the effect is directionally consistent ($-3\%$ to $-8\%$, or $-13\%$
under strict-log) across all other specifications.

\input{tables/tab_supplementary}

%% file: tables/tab_main_did.tex
\begin{table}[t]
\centering
\caption{Main diff-in-diff estimates: effect of AI tool adoption on architectural smell density (log ASD).}
\label{tab:main_did}
\begin{tabular}{lcc}
\toprule
 & Borusyak & TWFE-SA \\
\midrule
ATT ($\hat{\beta}$) & -0.0698*** & -0.0681*** \\
 & (0.0239) & (0.0238) \\
Effect (\%) & -6.7\% & -6.6\% \\
\bottomrule
\multicolumn{3}{l}{\footnotesize $^{*}p<0.1$; $^{**}p<0.05$; $^{***}p<0.01$} \\
\multicolumn{3}{l}{\footnotesize Both: 1,811 obs., 151 repos, repo and time FE, SEs clustered by repo.} \\
\end{tabular}
\end{table}

%% file: tables/tab_supplementary.tex
\begin{table}[t]
\centering
\caption{Supplementary analyses: denominator decomposition, inference robustness, attrition bounds, and sample sensitivity.}
\label{tab:supplementary}
\resizebox{\linewidth}{!}{%
\begin{tabular}{lccccc}
\toprule
Analysis & ATT ($\hat{\beta}$) & SE & 95\% CI & $p$-value & Effect (\%) \\
\midrule
\multicolumn{6}{l}{\textit{Denominator decomposition}} \\
Raw smell count & 0.0108 & 0.0486 & [-0.0845, 0.1061] & 0.8246 & 1.1\% \\
LOC growth (log KLOC) & 0.1201*** & 0.0401 & [0.0415, 0.1986] & 0.0027 & 12.8\% \\
Smells per package & -0.0318* & 0.0171 & [-0.0653, 0.0018] & 0.0636 & -3.1\% \\
\midrule
\multicolumn{6}{l}{\textit{Inference robustness}} \\
Wild cluster bootstrap & -0.0615*** & 0.0249 & [-0.1098, -0.0129] & 0.0097 & -6.0\% \\
\midrule
\multicolumn{6}{l}{\textit{Attrition bounds (Lee 2009)}} \\
Upper bound (pessimistic) & -0.0229 & 0.0195 & [-0.0611, 0.0152] & 0.2387 & -2.3\% \\
Lower bound (optimistic) & -0.1046*** & 0.0192 & [-0.1422, -0.0671] & 0.0000 & -9.9\% \\
\midrule
\multicolumn{6}{l}{\textit{Sample sensitivity}} \\
Excl.\ stale ($\geq$3 zero-mo, N=94) & -0.0413 & 0.0339 & [-0.1078, 0.0252] & 0.2234 & -4.0\% \\
Excl.\ stale ($\geq$6 zero-mo, N=128) & -0.0556** & 0.0251 & [-0.1048, -0.0063] & 0.0269 & -5.4\% \\
Packages $\geq$ 30 (N=109) & -0.0590*** & 0.0216 & [-0.1013, -0.0167] & 0.0063 & -5.7\% \\
Balanced ASD (equal type weight) & -0.0469*** & 0.0139 & [-0.0742, -0.0196] & 0.0007 & -4.6\% \\
Excl.\ outliers (KLOC $\geq$ 1) & -0.0791*** & 0.0188 & [-0.1159, -0.0423] & 0.0000 & -7.6\% \\
\bottomrule
\multicolumn{6}{l}{\footnotesize $^{*}p<0.1$; $^{**}p<0.05$; $^{***}p<0.01$. Primary ASD estimate: $\hat{\beta}$=-0.070, $p$=0.004.} \\
\end{tabular}%
}
\end{table}

%% file: sections/05_discussion.tex
\section{Discussion}\label{sec:discussion}

\subsection{Code Growth Without Proportional Architectural Degradation}\label{sec:disc_neutral}

Architectural smells are topological: cyclic dependencies require
mutual package imports; hub-like dependencies require excessive
fan-in and fan-out. Neither increases despite growth in code volume
($+12.8\%$) and package count ($+9.7\%$, $p = 0.032$), suggesting
new modules integrate without creating structural violations. UD's suggestive count increase
($+5.8\%$, raw $p = 0.032$; Holm-adjusted $p = 0.128$) would, if
confirmed by replication, be consistent with AI tools preferentially
adding imports to less-stable packages~\cite{martin2003agile}. It
does not survive multiplicity correction.

The limited warning-smell overlap (33.8\% non-co-occurring~\cite{esposito2025correlation})
offers a candidate, untested reconciliation with He et al.~\cite{he2026speed}:
AI tools may write more complex code within modules while respecting
between-module boundaries.

\subsection{A Methodological Lesson for Density-Normalized Mining Outcomes}\label{sec:disc_decomp}

Without decomposition, the $-6.7\%$ result would invite the
misleading headline ``AI tools improve architecture.'' The correct
interpretation is that adopters grow code volume faster than matched
controls without a proportional increase in architectural violations. This
generalizes: any density-normalized metric (defects/KLOC,
vulnerabilities/KLOC, churn/KLOC) where treatment plausibly affects
KLOC requires explicit decomposition. He et
al.'s~\cite{he2026speed} raw-count SonarQube metrics show warning
growth ($+30\%$) exceeding LOC growth ($+28.6\%$), indicating
per-line degradation at the code level. Smells are topological yet
ASD normalizes by KLOC (volumetric): the smells-per-package
alternative normalization (a denominator change: $-3.1\%$,
$p = 0.064$) shows a smaller, marginal effect consistent with the
denominator explanation.

\subsection{Implications}\label{sec:disc_implications}

\noindent\textbf{For practitioners.} These findings do not support
escalating architectural safeguards in response to agentic AI adoption
in the short term, but they warrant three monitoring practices:
(1) track raw smell counts alongside ASD, since density alone will
look improving even when counts are stagnant;
(2) treat 6-month windows as preliminary and re-examine at 12+ months
given the trajectory at $h = 6$ shows no inflection; and
(3) monitor coupling metrics ($Ce$, $I$) alongside smell counts as a
complementary signal.

\noindent\textbf{For researchers.} Density-normalized outcomes
require numerator/denominator decomposition under any treatment
plausibly affecting system size. The highest-value extensions are
greenfield projects (where AI influences architecture from
inception), longer observation windows, and dependency-graph
topology diffs to identify AI-introduced edges.

%% file: sections/06_threats.tex
\section{Threats to Validity}\label{sec:threats}

\textbf{Internal Validity.}
\textit{Differential attrition} is the most serious concern.
From the matched 90+90, 16/90 treatment (17.8\%) and 13/90 control
(14.4\%) were lost (5 controls dropped pre-Arcan for insufficient git
history; 24 repositories, 16 treatment and 8 control, lost during Arcan
analysis to JVM memory limits, concentrated among the largest projects). Lee bounds bracket the effect at
$[-9.9\%, -2.3\%]$: both bounds are negative, but the upper bound's
95\% CI includes zero ($p = 0.239$). Under worst-case attrition, the
effect cannot be distinguished from null.
\textit{Parallel trends} are supported by the event study Wald test
($p = 0.896$), though this cannot exclude smooth pre-treatment effects
absorbed by repository fixed effects.
\textit{Estimator sensitivity:} Borusyak and TWFE-SA agree closely
($-6.7\%$/$-6.6\%$). The Bacon decomposition (Sec.~\ref{sec:res_robustness}) confirms 89.1\% of
weight comes from clean comparisons, the bias Borusyak avoids by construction.
\textit{Stale observations:} 19.3\% of consecutive month-pairs share
identical commits; aggressive filtering ($N = 94$) attenuates the
effect to non-significance, though moderate filtering ($N = 128$,
$p = 0.027$) retains it.
\textit{Interference:} we assume no spillover between units (SUTVA):
shared libraries or fork relationships could violate this assumption,
though any spillover would attenuate estimates toward zero.
\textit{Co-adoption and selection:} the treatment may capture a
modernization bundle. Repository fixed effects and matching mitigate
time-invariant confounders, but unobserved time-varying changes
could confound the estimate.

\textbf{External Validity.}
We focus on mature Java OSS systems to isolate architectural effects
where structure is already established; findings apply to short-term
effects under agentic AI usage. The effect may differ in greenfield
projects where AI influences architecture from inception, or under
strict governance constraining AI-generated changes. Generalization is
further limited by language (Java, an Arcan constraint).
Survivorship is a further threat: repositories abandoned before
sampling are invisible, potentially biasing toward neutrality.

\textbf{Construct Validity.}
\textit{AI adoption proxy:} our treatment captures observable agentic
AI usage through configuration artifacts and commit metadata. Treated
repositories are a purposive subset, restricting external validity.
While this excludes autocomplete-only usage (false negatives) and may
include one-off experiments (false positives), prior work shows such
artifacts encode project-specific directives~\cite{jiang2026beyond}.
We validate this proxy empirically (Sec.~\ref{sec:meth_sample}):
temporal alignment and cross-signal consistency support its
reliability. Excluding single-event repos retains significance
($-6.8\%$, $p = 0.020$).
\textit{Control contamination:} 8/90 matched controls (8.9\%) showed
AI markers during their study windows. The corrected effect is
$-7.4\%$ (Sec.~\ref{sec:meth_sample}), and the qualitative pattern
(significant ASD decline, unchanged raw counts, large LOC growth)
is unchanged.
\textit{Arcan and architecture operationalization:} Arcan precision
is reported in the $70$--$100\%$ range~\cite{fontana2017arcan,sas2023atd}.
The trial edition was used. Arcan detects smells at class and package level:
our primary ASD aggregates both, with 57\% of cyclic dependencies
at class level. A container-only sensitivity ($-3.0\%$,
$p = 0.074$) is directionally consistent but non-significant. We
operationalize architecture through dependency-graph structure,
which does not capture design intent, conformance, or semantic
violations.
\textit{ASD normalization:} the ratio metric conflates numerator and
denominator changes. This is addressed via explicit decomposition
(Sec.~\ref{sec:res_decomposition}) and alternative normalization
(smells per package, $p = 0.064$).
\textit{Reproducibility:} mitigated by detection-pattern disclosure
(Sec.~\ref{sec:meth_sample}), pinned tool versions, fixed random
seeds, and a complete replication package.

%% file: sections/07_conclusion.tex
\section{Conclusion}\label{sec:conclusion}

This study presented the first causal mining study of agentic AI
adoption's effect on software architecture, analyzing 1{,}811
monthly snapshots across 151 Java repositories with a staggered
difference-in-differences design. The apparent 6.7\% ASD reduction
is a composition effect: repositories with observable agentic AI
adoption exhibit increased code volume without proportionally more
architectural smells. We find no evidence that the observed agentic
AI tools degrade software architecture in established projects over
a six-month window, though under worst-case attrition the
confidence interval includes zero.

Future work should extend the observation window and study
greenfield projects. Dependency-graph topology diffs are needed to
confirm the mechanism. The denominator-decomposition lesson
generalizes to any density-normalized mining outcome under
size-affecting treatment, and we encourage empirical AI-tool studies
to adopt this practice by default.